\documentclass[preprint,amssymb,showpacs,floatfix]{revtex4}
\usepackage{graphicx}
\newcommand{\bea}{\begin{eqnarray}}
\newcommand{\eea}{\end{eqnarray}}
\newcommand{\be}{\begin{equation}}
\newcommand{\ee}{\end{equation}}
\newcommand{\rmscr}[1]{{\hbox{\scriptsize \rm{#1}}}}
\newcommand{\rmmat}[1]{{\hbox{\rm{#1}}}}
\def\apj{Astrophys. J.}
\def\apjl{Astrophys. J. Lett.}
\def\apjs{Astrophys. J. Suppl. Ser.}
\def\aap{Astron. Astrophys.}
\def\apss{Astrophys. Space Sci.}
\def\araa{Annu. Rev. Astron. Astrophys.}
\def\physrep{Phys. Rep. }
\def\mnras{Mon. Not. Roy. Astron. Soc.}
\def\pasj{Pub. Astron. Soc. Japan}
\def\ie{{\em i.e.\ }}
\def\eg{{\em e.g.\ }}
\def\etal{{\em et al.\ }}

\def\rmmat#1{{\hbox{\rm #1}}}
\def\rmscr#1{\rmmat{\scriptsize #1}}
\begin{document}

\title{Electron Positron Capture Rates and the Steady State Equilibrium
Condition for Electron-Positron Plasma with Nucleons}
\author{Ye-Fei Yuan} 
\email{yfyuan@ustc.edu.cn}
\affiliation{Center for Astrophysics, University of Science and Technology
	of China, Hefei, Anhui 230026, P.R. China}

\begin{abstract}
The reaction rates of the beta processes for all particles 
at arbitrary degeneracy are derived, and
an {\it analytic} steady state equilibrium condition $\mu_n=\mu_p+2\mu_e$
which results from the equality of electron and positron capture rates
in the hot electron-positron plasma with nucleons is also found,
if the matter is transparent to neutrinos.
This simple analytic formula is valid only if electrons are
nondegenerate or mildly degenerate, which is generally satisfied in the hot
electron-positron plasma. Therefore, it can be used to efficiently 
determine the steady state of the hot matter with plenty of positrons.
Based on this analytic condition, given the baryon number density and the
temperature, if the nucleons are nondegenerate, only one algebraic 
equation for determining the electron fraction is obtained, 
which shows the great advantage of the analytic equilibrium condition.

\end{abstract}

\maketitle

\section{Introduction}
Gamma Ray Bursts (GRBs) \citep{2002ARA&A..40..137M} and 
core collapse supernovae(SNe) \citep{2001pnsi.conf..333J}
are two of the most violent events in our universe. Ironically,
their explosion mechanisms are still mysterious. 
Recently, the central engine of GRBs is believed to be related to 
the hyperaccretion of a stellar-mass black hole at extremely high rates
from $\sim$ 0.01 to 10 $M_{\odot}$s$^{-1}$ 
\citep{1992ApJ...395L..83N,2003ApJ...586.1254P,2003ApJ...588..931B}.
In such an accretion disk,
matter is so dense that photons are trapped. The possible 
channel for energy release is either neutrino emission 
which is mainly from the electron-positron ($e^{\pm}$) capture on nucleons and 
$e^{\pm}$ annihilation, or outflows from the disk.
Whatever a successful central engine is, it ejects a hot
fireball which consists of the radiation field 
($e^{\pm}$/photons) 
and baryons. 
The ratio of neutrons to protons, or equivalently, 
the electron fraction, is crucial to the observed radiation 
from GRBs \citep{1999ApJ...521..640D,2002ApJ...573..770P}, its
dynamic evolution \citep{2000PhRvL..85.2673F,2003ApJ...588..931B} and 
the nucleosynthesis in the disk or fireball
\citep{2003ApJ...586.1254P,2003ApJ...588..931B}.
For instance, the inelastic collisions between 
neutrons and protons produce observable multi-GeV neutrino emission
\citep{2000PhRvL..85.1362B,2002ARA&A..40..137M}. In addition,
the two component fluid of neutrons and protons 
significantly changes the fireball interaction with an external medium
which is supposed to produce the observed electromagnetic radiation
from GRBs and their afterglows \citep{2003ApJ...585L..19B}; and
the electron fraction $Y_e$ strongly affects the equation of state of 
the hyperaccretion disk and the neutrino emissions from it.

Roughly speaking, SNe are powered by the iron core collapse of their 
progenitors. Most numerical simulations have shown not only the failure of 
the prompt shock, but the failure of its revival by the delayed
neutrino emission from the protoneutron star (PNS). The result (explosion 
or not) sensitively depends on the input microphysics, such as the 
electron capture, the neutrino emission, neutrino-matter interactions,
the equation of state, rotation, magnetic field, general relativity effects, 
and so on (see ref.\citep{2001pnsi.conf..333J}, and 
references therein). Without a doubt, 
weak interactions, especially $e^{\pm}$ captures and neutron 
decay, play a key role in both GRBs and SNe. During the accretion 
or collapse, these processes exhaust electrons, thus decrease the degenerate
pressure of electrons. Meanwhile, they produce neutrinos  which carry 
the binding energy away to power the explosions. 
Therefore, electron capture is crucial
to the formation of the bounce shock of SNe, and the resulting neutrino spectra
strongly influence the neutrino-matter interactions which are energy dependent
and are essential for collapsing simulations \citep{2003RvMP...75..819L}.

The existence of a hot state with nucleons is
the common characteristics of both GRBs and SNe, as well as
the PNS, the bounce shock and the early universe
\citep{2002RvMP...74.1015W}. In these systems,
the beta reactions  
are the most important physical processes \citep{Pina64}. 
If the system is transparent to neutrinos, 
the steady state is achieved via 
the following beta reactions \citep{1967SvA....10..970I},
\bea
e^-+p &\rightarrow & n+ \nu_e, \label{eq:e_cap}\\
e^++n &\rightarrow & p+ \bar{\nu}_e, \label{eq:p_cap}\\
n     &\rightarrow & p+e^- + \bar{\nu}_e .  \label{eq:n_dec}
\eea
These beta reaction rates are calculated in 
the previous studies, usually under one of three approximations: 
the nondegenerate approximation \citep{Pina64}, 
the degenerate approximation 
\citep{1979ApJ...232..541F,1983bhwd.book.....S}, 
and the elastic approximation
in which there is no energy transfer to nucleons 
\citep{1985ApJS...58..771B}. 
In this paper, applying the structure function
formalism developed by \citet{1998PhRvD..58a3009R}(see also 
\citep{1998PhRvC..58..554B}), I derive the reaction 
rates of the beta processes in the dense subnuclear matter
for all particles at arbitrary degeneracy.
In addition, I find an analytic expression for determining
the kinetic equilibrium between electron capture and positron
capture, which is efficient to determine the steady state of the hot 
matter with plenty of positrons. 
If the neutrinos are partially trapped, neutrino transport should 
be considered carefully, which is out of the scope of this work
\citep{2003PhRvD..68f3001S}.

\section{Reaction rates.}
\begin{figure}
\includegraphics[width=6.0in]{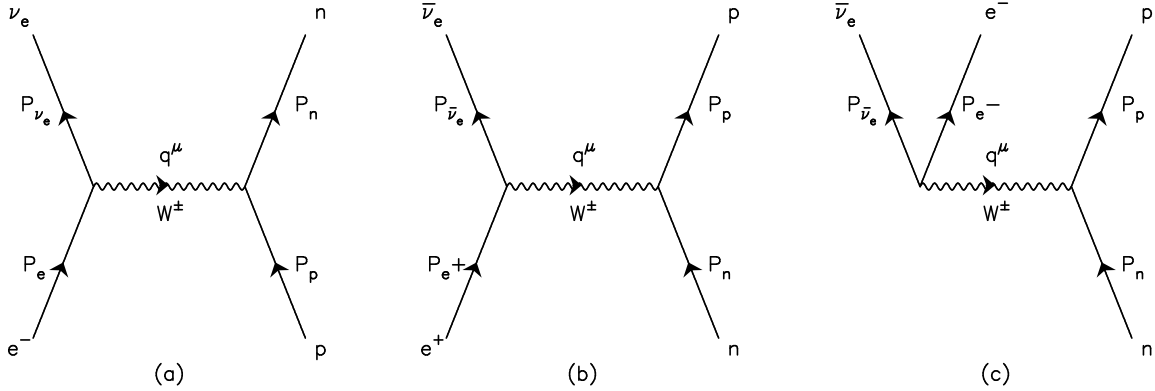}
\caption{
The lowest order Feynman diagrams for $\beta$ processes.
(a) Electron capture by proton: $e^- + p \rightarrow n + \nu_{e}$,
    the energy conservation requires $E_e=q_0+E_{\nu}$.
(b) Positron capture by neutron: $e^+ + n \rightarrow p + \bar{\nu_{e}}$,
    the energy conservation requires $E_{e^+}=q_0+E_{\bar{\nu}}$.
(c) Neutron decay: $n \rightarrow p + e^- + \bar{\nu_{e}}$,
    the energy conservation requires $E_{e}=-q_0-E_{\bar{\nu}}$.
\label{fig:feynman}
}
\end{figure}

The lowest order Feynman diagrams for reactions 
(\ref{eq:e_cap})-(\ref{eq:n_dec}) 
are shown in Fig.~\ref{fig:feynman}. Because in our consideration
the energies of leptons are less than a few hundred MeV which is 
greatly less than the rest mass of $W^{\pm}$ mesons, thus the
interaction Lagrangian from Weinberg-Salam theory is reduced to 
the original Fermi's current-current interaction form. 
From Fermi's golden rule, the reaction rates of 
the processes (\ref{eq:e_cap})-(\ref{eq:n_dec}) read
(we set $\hbar=c=k_{\rm B}=1$)
\be
\lambda=2\int \prod_{i=1}^{4} \left[\frac{d^3\vec{p_i}}{(2\pi)^3} 
  \right](2\pi)^4
  \delta^{(4)}(P_i-P_f) |M|^2 \cal{F},  \label{eq:rate_g}
\ee
where $P_i=(E_i,\vec{p_i})$ denotes the four-momentum of
particle $i$ ($i=\nu_e/\bar{\nu}_e, e^-/e^+, n, p$), 
$p_i=|\vec{p_i}|$, and $P_i$ and 
$P_f$ are the total initial and final momentum, respectively.
$\cal{F}$ denotes the final--states blocking factor.
In reaction (1)-(3), $\cal{F}$=$f_e f_p (1-f_n)$, 
 $\cal{F}$=$f_{e^+} f_n (1-f_p)$ ,  $\cal{F}$=$f_n (1-f_p) (1-f_e) $ 
respectively, where $f_i$ is the Fermi-Dirac function of
particle $i$. In this paper, we just consider the case 
that the emitted neutrinos can escape freely from the system.
$|M|^2$ in the above equation is the averaged transition rate which 
depends on the initial and final states of all participating
particles. 
Screening corrections to the electron
capture rates in dense astrophysical environments have been 
investigated by many authors
\citep{1974ApJ...194..385C}. 
In the hot and dense $npe^{\pm}$ gas, the Debye radius $r_{\rm D}$ 
is about 3\,fm$(T/10^{10}{\rm K})^{1/2}(n_{\rm B}/0.01{\rm fm}^{-3})^{-1/2}(Y_e/0.5)^{-1/2}$, 
where $n_{\rmscr{B}}=n_n(\mu_n,T)+n_p(\mu_p,T)$
is the baryon number density.  Thus in our calculations, 
Coulomb waves are approximately replaced 
with plane waves for charged particles.
As argued in \citet{1998PhRvD..58a3009R}, 
for nonrelativistic noninteracting baryons 
($n_{\rm B}< 5n_{\rm nuc}$), where $n_{\rm nuc}=0.16$\,fm$^{-3}$
is the empirical nuclear equilibrium density, the transition
rate averaged over the initial spins becomes a constant, \ie
$|M|^2=G_{\rmscr{F}}^2\cos^2\theta_{\rmscr{C}}(1+3g_{\rmscr{A}}^2)$, 
here $G_{\rmscr{F}} \simeq 1.436 \times 10^{-49}~\rmmat{erg}~ 
\rmmat{cm}^{3}$ is the Fermi weak interaction constant,
$\theta_{\rmscr{C}}$ $(\sin\theta_{\rmscr{C}}=0.231)$ 
is the Cabibbo angle, and 
$g_{\rmscr{A}}=1.26$ is the axial-vector coupling constant.
For the case of relativistic interacting baryons, 
the transition rate is expressed in terms of 
the target particle retarded polarization tensor in the
relativistic mean field theory for interacting baryons,
the resulting expression is so complicated (for more detail, 
see \citep{Horowitz:1990it,1998PhRvD..58a3009R}).
In this paper, the beta reactions at subnuclear density 
are considered, therefore, Taking $|M|^2$ to be constant 
is a good approximation.

Using the structure function formalism developed by 
\citet{1998PhRvD..58a3009R}, 
the above integrations can be simplified into only three dimensional ones,
therefore, the rate of the $e^{\pm}$-captures $\lambda_{e^-p}, \lambda_{e^+n}$,
and the rate of neutron decay $\lambda_{n}$ are given by
\bea
\lambda_{e^-p}&=&\frac{1}{8\pi^4}|M|^2 
	\int_0^{\infty}d E_{\nu} 
 	\int_{m_e-E_{\nu}}^{\infty} dq_0  \nonumber \\
	& \times &\int_{|p_e-p_{\nu}|}^{|p_e+p_{\nu}|}dq 
	E_{\nu} E_e f_eS_{p \rightarrow n}(q_0,q) q ,
		\label{eq:rate_e_cap1} \\
\lambda_{e^+n}&=&\frac{1}{8\pi^4}|M|^2 
	\int_0^{\infty}d E_{\bar{\nu}} 
	\int_{m_e-E_{\bar{\nu}}}^{\infty} dq_0 \nonumber \\
	& \times &\int_{|p_{e^+}-p_{\bar{\nu}}|}^{|p_{e^+}+p_{\bar{\nu}}|}dq 
	E_{\bar{\nu}} E_{e^+} f_{e^+}S_{n \rightarrow p}(q_0,q) q,
		\label{eq:rate_p_cap1}\\
\lambda_{n}&=&\frac{1}{8\pi^4}|M|^2 
	\int_0^{\infty}d E_{\bar{\nu}} 
 	\int^{-(m_e+E_{\bar{\nu}})}_{-\infty} dq_0   \nonumber \\
	& \times &\int_{|p_e-p_{\bar{\nu}}|}^{|p_e+p_{\bar{\nu}}|}dq 
	E_{\bar{\nu}} E_e (1-f_e)S_{n \rightarrow p}(q_0,q) q,
		\label{eq:rate_n_dec1}
\eea
where $S_{i\rightarrow j}(q_0,q)$ is the so-called dynamic 
form factor or structure function
which characterizes the isospin response of the system 
\citep{1998PhRvD..58a3009R}, 
and $q_0=E_f-E_i$, $q=|\vec{q}|=|\vec{p_f} - \vec{p_i}|$
denote the momentum and energy transfer.

The expression
$S_{i\rightarrow j}(q_0,q)$ is given by
\be
S_{i\rightarrow j}(q_0,q)=
	\frac{m_i m_f T}{\pi q}
	\frac{z+\xi_-}{1-\exp(-z)} 
 \label{eq:form} , 
\ee
where
\bea
z&=&\frac{q_0+\mu_i-\mu_j}{T}, \\
\xi_-&=&\ln\left[\frac{1+\exp((E_-^i-\mu_i)/T)}{1+\exp(E_-^i+q_0-\mu_j)} 
\right], \\
E_-^i&=m_i+&\frac{m_j^2(q_0+m_i-m_j-q^2/2m_j)^2}{2m_iq^2},
	\label{eq:struct}
\eea
where $\mu_i$ and  $m_i$ are the chemical potential and the mass of 
baryons.  
In Eqs.~(\ref{eq:rate_e_cap1})-(\ref{eq:rate_n_dec1}),
$E_e=q_0+E_{\nu}$,
$E_{e^+}=q_0+E_{\bar{\nu}}$, and 
$E_{e}=-q_0-E_{\bar{\nu}}$ (see Fig. 1). 
Equations~(\ref{eq:rate_e_cap1})-(\ref{eq:rate_n_dec1}) are valid for
nonrelativistic and noninteracting baryons \citep{1998PhRvD..58a3009R}.
Below the nuclear density, this is a good approximation.
It should be emphasized that Eqs.~(\ref{eq:form})-(\ref{eq:struct}) 
differ from the corresponding equations in \citet{1998PhRvD..58a3009R} in 
which the mass difference between nucleons is neglected. However, in
order to investigate the rate of neutron decay, it is necessary to keep 
the mass difference in Eqs.~(\ref{eq:form})-(\ref{eq:struct}).

Analogous to the analysis in \citet{1998PhRvD..58a3009R}, 
it is easy to obtain the previous results
in the nondegenerate and degenerate limits of baryons.
For instance, the results in the nondegenerate limit corresponding to 
Eqs.~(\ref{eq:rate_e_cap1})-(\ref{eq:rate_n_dec1}) 
are shown below,
\bea
\lambda_{e^-p}&\simeq&\frac{1}{2\pi^3}|M|^2 n_p\int_Q^{\infty}dE_e
        E_ep_e(E_e-Q)^2f_e, 
        \label{eq:rate_e_cap2} \\
\lambda_{e^+n}&\simeq&\frac{1}{2\pi^3}|M|^2 n_n\int_{m_e}^{\infty}dE_e
        E_ep_e(E_e+Q)^2f_{e^+},
        \label{eq:rate_p_cap2} \\
\lambda_{n}&\simeq&\frac{1}{2\pi^3}|M|^2 n_n\int_{m_e}^QdE_e
        E_ep_e(Q-E_e)^2  \nonumber \\
        && \times (1-f_{e}), 
        \label{eq:rate_n_dec2} 
\eea
where $Q=m_n-m_p$ is the mass difference between neutron and proton,
$n_i=2(m_iT/2\pi)^{3/2}\exp({\eta_i})$ is the number density of neutrons
and protons in the nondegenerate limit, and $\eta_i=(\mu_i-m_i)/T$ is the
reduced chemical potential. 
The above approximate rates are frequently cited 
in the literature to discuss the 
kinetic equilibrium for the $\beta$-processes and 
the emissivity of neutrino emission, 
even though its validity should be checked carefully
\cite{1967SvA....10..970I,2003ApJ...586.1254P,2003ApJ...588..931B}.

At $m_e,\mu_e,Q<T$,
the above reaction rates can be simplified further into
\bea
\lambda_{e^-p}&\simeq&2^{-1.5}\pi^{-4.5}|M|^2m_p^{1.5}T^{6.5} \nonumber \\
        && \times \exp({\eta_p+\mu_e/T-Q/T)} \nonumber \\
        && \times [I(4)+2I(3)Q/T] ,
        \label{eq:rate_e_cap3} \\
\lambda_{e^+n}&\simeq&2^{-1.5}\pi^{-4.5}|M|^2m_n^{1.5}T^{6.5} \nonumber \\
        && \times \exp({\eta_n-\mu_e/T}) \nonumber \\
        && \times [I(4)+2I(3)Q/T] ,
        \label{eq:rate_p_cap3}\\
\lambda_{n}&\simeq&1.63\times 2^{-1.5}\pi^{-4.5}|M|^2m_n^{1.5}T^{1.5}
                m_e^5 ,
        \label{eq:rate_n_dec3}
\eea
here $I(3)=7\pi^4/120$, and $I(4)=45\zeta(5)/2$, $\zeta(5)=1.037$ is
the Riemann $\zeta$-function.  The definition of $I(n)$ 
is $ I(n)\equiv\int_0^{\infty}(e^x+1)^{-1}x^ndx$. This
integral can be found in \citet{1965tisp.book.....G}.

\begin{figure}
\includegraphics[width=3.0in]{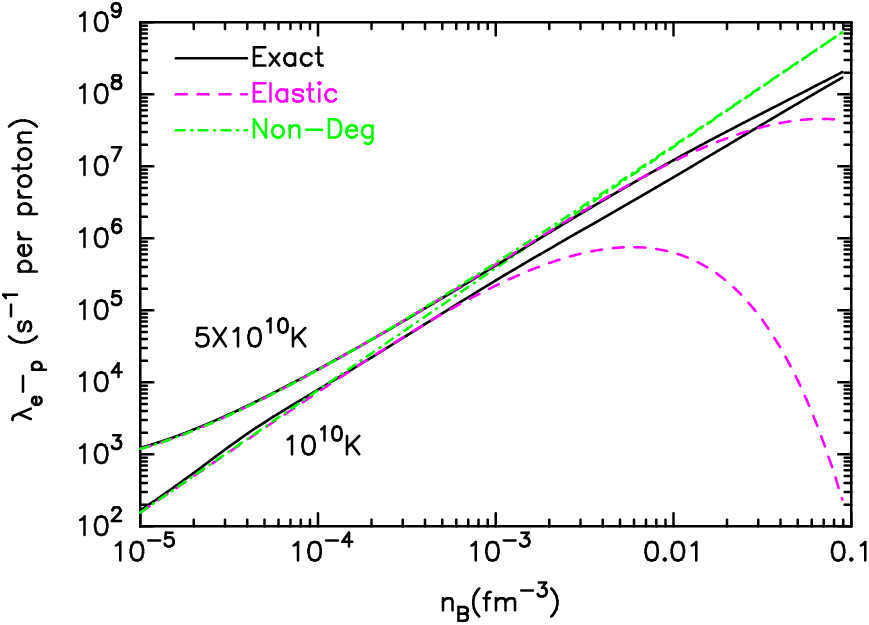}
\caption{Electron capture rates on protons in the dense subnuclear
 matter out of the chemical equilibrium as a function of
 the baryon number density $n_{\rmscr{B}}$ at different temperatures.
 The electron fraction $Y_e$ is taken to be 0.3. The solid curves 
 show the results from Eq.~(\ref{eq:rate_e_cap1}), the dashed 
 curves are the elastic results from Eq.~(\ref{eq:elastic}), 
 the dot-dashed curves are the nondegenerate results from 
 Eq.~(\ref{eq:rate_e_cap2}).
\label{fig:e_cap}
}
\end{figure}

In the elastic limit, 
$n_p$ in Eq.~(\ref{eq:rate_e_cap2}) is replaced by
\be
\eta_{pn}=(n_p-n_n)/(1-e^{(\eta_n-\eta_p)/T}). \label{eq:elastic}
\ee
It is generally believed that electron capture rate under 
the elastic approximation in some sense introduces the effects of the 
degeneracy of baryons, 
so it should be more accurate than that under the nondegenerate 
approximation. Suppose that the nuclei are dissolved completely 
into nucleons at high temperature. 
Figure~\ref{fig:e_cap} shows the differences between our
electron capture rate in the dense subnuclear matter out of
the chemical equilibrium 
and the previous approximate results. The electron fraction 
$Y_e=(n_{e^-}-n_{e^+})/n_{\rmscr{B}}$ is assumed to be 0.3. 
The number density of particles at any degeneracy is expressed
in terms of the Fermi-Dirac functions \citep{1998ApJS..117..627A}.
The multidimensional integrations and the Fermi-Dirac functions
are calculated using the mixture of Gauss-Legendre and Gauss-Laguerre
quadratures \citep{1998ApJS..117..627A}.
From Fig.~\ref{fig:e_cap}, it is evident that there are great 
differences between the elastic 
results and the general results (Eq.~\ref{eq:rate_e_cap1}) 
when baryons become degenerate.
As shown in Eq.~(\ref{eq:elastic}), 
the capture rates under the elastic approximation always decrease 
exponentially when nucleons become degenerate, which is 
qualitatively correct in the dense nuclear matter near the
$\beta$-equilibrium. 
However, this conclusion is not correct obviously in the dense subnuclear
matter out of the chemical equilibrium. 
The elastic approximation is also good
when $\mu_p \simeq \mu_n$ because the favored energy 
transfer is zero. 
In short, the elastic approximation underestimates the electron capture rate,
when baryons become degenerate.

\section{Conditions for kinetic equilibrium.}
\subsection{Chemical equilibrium condition for neutrino trapping.}
If neutrinos are trapped in a system, as what happens 
in the interior of a PNS, 
the inverse reactions corresponding to beta processes
(\ref{eq:e_cap})-(\ref{eq:n_dec}) by absorption of neutrinos  
can take place:
\bea
e^-+p &\rightleftharpoons & n+ \nu_e, \label{eq:e_capR}\\
e^++n &\rightleftharpoons & p+ \bar{\nu}_e, \label{eq:p_capR}\\
n     &\rightleftharpoons & p+e^- + \bar{\nu}_e  , \label{eq:n_decR}
\eea
and their rates are equal to these of the corresponding
beta reactions when the system reaches the chemical equilibrium.
Because both the photons and neutrinos are trapped, 
the chemical equilibriums
$\gamma + \gamma \rightleftharpoons e^+ + e^- 
\rightleftharpoons \nu_{e} + \bar{\nu}_{e} $ give($\mu_{\gamma}=0$), 
\bea
\mu_{e^+} &=& -\mu_{e^-}, \\
\mu_{\bar{\nu}_{e}} &=& -\mu_{\nu_e}.
\eea
Based on the standard arguments in the theory of thermodynamics, the 
chemical equilibrium of any one
of the above reversible reactions (\ref{eq:e_capR})-(\ref{eq:n_decR})
gives the same chemical equilibrium condition 
for both cold $npe^-$ and hot $npe^{\pm}$ gases,
\be
\mu_n+\mu_{\nu_e}=\mu_p+\mu_e.
	\label{eq:trapping1}
\ee

If the chemical potential of the trapped neutrinos 
is zero, the above equation reduces to
\be
\mu_n = \mu_p+\mu_e,
	\label{eq:trapping2}
\ee
%
and the corresponding number density of the trapped neutrinos is
written as
\be
n_{\nu_e} = n_{\bar{\nu}_e} = \frac{T^3}{2 \pi^2} \int_0^{\infty} 
	\frac{x^2 dx}{e^x +1} = \frac{3T^3}{4 \pi^2} \zeta (3),
	\label{eq:trappingN}
\ee
where $\zeta (3)\simeq 1.202$.
It is clearly shown in Eq.~(\ref{eq:trappingN}) that the number 
density of the trapped neutrino is not zero at all when their
chemical potential is zero. Only when $T \rightarrow 0$, 
$n_{\nu_e} = n_{\bar{\nu}_e} \rightarrow 0$. Therefore, Eq.~(\ref{eq:trapping2})
can also be understood as the chemical equilibrium condition for
cold $npe^-$ gas under $\beta$-equilibrium when neutrinos can escape 
freely from the system.

\subsection{Kinetic equilibrium condition for neutrino escaping.}
In principle, if the matter is transparent to neutrinos which carry
energy away, the equilibrium of the reactions 
(\ref{eq:e_cap})-(\ref{eq:n_dec}) can not be treated as a chemical 
equilibrium problem\citep{1983bhwd.book.....S}. 
Suppose that the dynamic time scale of the system under consideration is 
greater than that of the reactions
(\ref{eq:e_cap})-(\ref{eq:n_dec}), the general condition for the 
$\beta$-equilibrium is given by
\bea
&&\lambda_{e^-p}(\mu_n,\mu_p,\mu_e,T) \nonumber \\
&&=\lambda_{e^+n}(\mu_n,\mu_p,\mu_e,T)
+\lambda_{n}(\mu_n,\mu_p,\mu_e,T). 
	\label{eq:equil}
\eea
Given $n_{\rm B}$ and $T$, there are two additional
trivial conditions. One is the conservation of the baryon number,
\be
n_n(\mu_n,T)+n_p(\mu_p,T) = n_{\rm B},
	\label{eq:baryon}
\ee
the other is the charge neutrality,
\be
n_{e^-}(\mu_{e},T)-n_{e^+}(\mu_{e},T)=n_{p}(\mu_{p},T).
	\label{eq:charge}
\ee
Consequently, the chemical potentials of neutron, proton and electron
($\mu_n, \mu_p, \mu_e$)
can be determined by a set of closed equations, \ie, 
Eqs.~(\ref{eq:equil})-(\ref{eq:charge}).

\section{Analytic $\beta$-equilibrium conditions for neutrino escaping.}
\subsection{$\beta$-equilibrium condition for cold $npe^-$ gas.}
For ideal, cold $npe^-$ gas, we can set the chemical potential of 
the neutrinos in Eq.~(\ref{eq:trapping1}) to zero because their number density 
is zero at $T=0$ (see Eq.~(\ref{eq:trappingN})), therefore, 
the chemical equilibrium conditions for the ideal, cold $npe^-$ gas
is Eq.~(\ref{eq:trapping2}), \ie, $\mu_n = \mu_p+\mu_e$  \citep{1983bhwd.book.....S}.

In the following, we will re-derive the chemical equilibrium condition
for the cold $npe^-$ gas Eq.~(\ref{eq:trapping2}) from the viewpoint of the 
equality of the reaction rates of the beta processes. In the interior
of an old neutron star, 
the electrons are degenerate ($E_e \gg T$), thus the number density of
positrons are exponentially depressed.
As the typical energy of emitted neutrinos is of order            
the temperature, the energy and the momentum 
of the emitted neutrinos can be neglected 
comparing to the Fermi energy of particles, \ie, $E_p+E_e=E_n$,
and $\vec{p_p}+\vec{p_e}=\vec{p_n}$.
So the kinetic equilibrium requires $\lambda_{e^-p}=\lambda_{n}$,
therefore,
\bea
0&=&\lambda_{e^-p}-\lambda_{n} \nonumber \\
 &\simeq& 2\int \prod_{i=1}^{4} \left[\frac{d^3\vec{p_i}}{(2\pi)^3}
  \right](2\pi)^4 |M|^2  
	\delta^{(4)}(P_e + P_p - P_n) \nonumber \\ 
 &&	[f_p(1-f_n)f_e -(1-f_p)f_n(1-f_e) ] \nonumber \\
 &\simeq& 2\int \prod_{i=1}^{4} \left[\frac{d^3\vec{p_i}}{(2\pi)^3}
  \right](2\pi)^4 |M|^2  \delta^{(4)}(P_e + P_p - P_n) \nonumber \\ 
 &&	f_p(1-f_n)(1-f_e) e^{-E_e} [e^{\mu_e/T}-e^{(\mu_n-\mu_p)/T}].
        \label{eq:app_d}
\eea
In the above derivations, we use the even property of the Dirac-delta function, 
\ie, $\delta^{(4)}(P_e + P_p - P_n)=\delta^{(4)}(P_n - P_e - P_p )$.
It is evident that Eq.~(\ref{eq:app_d}) eventually results in the 
well-known chemical equilibrium condition 
Eq.~(\ref{eq:trapping2}) which is generally used to determine the equation
of state of the dense matter in the interior of an old neutron star.

\subsection{Analytic steady state equilibrium condition for 
hot $npe^{\pm}$ gas.}
It is well known that in a system with plenty of positrons,
the $e^{\pm}$ pair must not be degenerate ($T>E_e$), if not, 
the number density of positrons will decrease exponentially,
therefore, $\exp((E_e-\mu_e)/T)\geq 1$.
Before we derive the analytic dynamical equilibrium condition for
such system from which the emitted neutrinos can escape 
freely, we first explain why the condition is not 
the well-known result of Eq.~(\ref{eq:trapping2}).
Suppose that Eq.~(\ref{eq:trapping2}) is satisfied, for the reason 
described in \S III.A., the chemical potential of the trapped neutrinos
is zero, thus the ratio of the number density of the trapped 
neutrinos to that of the electrons is about $T^3/E_e^3 > 1$, if
the electrons are not degenerate! This conclusion is obviously 
contradictory to the precondition that the system is transparent to
neutrinos.

In such system, comparing to the rate of positron capture by neutrons,
the rate of neutron decay
could be neglected before reaching the degenerate limit.
Thus the $\beta$-equilibrium condition for hot $npe^{\pm}$ gas
is \citep{2003ApJ...588..931B,2003ApJ...586.1254P}
\be
\lambda_{e^-p}=\lambda_{e^+n},
\label{eq:equil_pair}
\ee
if the hot electron-positron plasma
with nucleon is transparent to neutrinos.
Another good approximation is the elastic approximation,
that is, the energy of the emitted neutrinos is of order that of the captured
$e^{\pm}$, \ie, $E_{\nu}\simeq E_e$, $E_{\bar{\nu}}\simeq
E_{e^+}$, and thus $E_n\simeq E_p$. 

Under these approximations, the kinetic equilibrium 
condition Eq.~(\ref{eq:equil_pair}) gives
\bea
0&=&\lambda_{e^-p}-\lambda_{e^+n} \nonumber \\
 &\simeq& 2\int \prod_{i=1}^{4} \left[\frac{d^3\vec{p_i}}{(2\pi)^3}
  \right](2\pi)^4 |M|^2
        \delta^{(4)}(P_e + P_p - P_n - P_{\nu_e}) \nonumber \\
	&& [f_p f_e (1-f_n) - (1-f_p) f_{e^+} f_n] \nonumber \\
 &\simeq& 2\int \prod_{i=1}^{4} \left[\frac{d^3\vec{p_i}}{(2\pi)^3}
  \right](2\pi)^4 |M|^2
        \delta^{(4)}(P_e + P_p - P_n - P_{\nu_e}) \nonumber \\
	&& f_p(1-f_n)f_{e^+}  \nonumber \\
        && \left[\frac{e^{(E_e+\mu_e)/T}+1}
          {e^{(E_e-\mu_e)/T}+1}
          -e^{((E_p-E_n)-(\mu_p-\mu_n))/T}\right] \nonumber \\
 &\simeq& 2\int \prod_{i=1}^{4} \left[\frac{d^3\vec{p_i}}{(2\pi)^3}
  \right](2\pi)^4 |M|^2
        \delta^{(4)}(P_e + P_p - P_n - P_{\nu_e}) \nonumber \\
	&& f_p(1-f_n)f_{e^+}[e^{2\mu_e/T} -e^{(\mu_n-\mu_p)/T}] .
	\label{eq:app_nd}
\eea
Therefore, from Eq.~(\ref{eq:app_nd}) we obtain the beta-equilibrium 
condition for the $e^{\pm}$ plasma with nucleons
\be
\mu_n=\mu_p+2\mu_e . \label{eq:nd_eq}
\ee
It should be emphasized that during the above derivation, it is
not assumed whether the baryons are degenerate or not.
On the other hand, Eq.~(\ref{eq:nd_eq}) is still valid 
under the degeneracy of baryons, which is neglected 
completely in the approximate reaction rates 
at the beginning. However, it should be pointed out that 
wherever baryons are degenerate in astrophysics, 
neutrinos are generally trapped and likely thermalized.
In these post-neutrino trapping environments, the analytic 
steady state equilibrium condition we derived here is
not valid, because the effects of neutrino trapping are
not included in our consideration.
It is not a surprise to notice that 
the analytic equilibrium condition Eq.~(\ref{eq:nd_eq}) 
can also be drawn in the completely nondegenerate limit.
Setting the equality of Eq.~(\ref{eq:rate_e_cap3}) and 
Eq.~(\ref{eq:rate_p_cap3}), the equilibrium condition 
is also obtained. In the following, we have another heuristic 
but not very strict derivation.
If all particles are nondegenerate, we have 
\cite{1983bhwd.book.....S,1985ApJS...58..771B}
\bea
\lambda_{e^-p} &\propto& 
           n_{e^-}n_p 
                \propto 
           \exp(\eta_p+\eta_e) 
	\label{eq:rate_e_cap4}, \\
\lambda_{e^+n} &\propto& 
           n_{e^+}n_n 
                \propto  
	 \exp(\eta_n-\eta_e) 
	\label{eq:rate_p_cap4} .
\eea
Setting the equality of Eq.~(\ref{eq:rate_e_cap4}) and 
Eq.~(\ref{eq:rate_p_cap4}), 
the analytic condition Eq.~(\ref{eq:nd_eq}) is obtained again.

The advantage of having the analytic equilibrium condition at hand
is obvious. For instance, we can derive some useful formula in
the nondegenerate limit of baryons. In such limit, using the 
Saha equation gives the ratio of neutron baryon number density 
to that of proton (\eg \cite{1982ApJ...252..741F}), 
\be
\frac{n_n}{n_p} \equiv 1-\frac{1}{Y_e}=e^{\frac{\mu_n-\mu_p-Q}{T}}
	=e^{\frac{2\mu_e-Q}{T}}.
        \label{eq:Ye1}
\ee
At the last step, the analytic equilibrium condition
Eq.~(\ref{eq:nd_eq}) is applied. On the other hand, for the
relativistic $e^{\pm}$, there exits an exact expression for
the electric charge density in terms of the chemical potential of electrons
(\eg \cite{1996ApJS..106..171B}),
\be
n_{e^-} - n_{e^+} \equiv n_{\rmscr{B}}Y_e
=\frac{T^3}{3 \pi^2} \left[\left(\frac{\mu_e}{T}\right)^3+\pi^2
\left(\frac{\mu_e}{T}\right)\right]
        \label{eq:Ye2}
\ee
Substituting Eq.~(\ref{eq:Ye1}) into Eq.~(\ref{eq:Ye2}),
we have only one simple equation to determine the electron chemical
potential $\mu_e$,
\be
\frac{T^3}{3 \pi^2} \left[\left(\frac{\mu_e}{T}\right)^3+\pi^2
\left(\frac{\mu_e}{T}\right)\right]-\frac{n_{\rmscr{B}}}{1+e^{(2\mu_e-Q)/T}}
=0.
\label{eq:app_mue}
\ee
After that, the electron fraction $Y_e$ can be directly calculated from
Eq.~(\ref{eq:Ye2}).

At $\mu_e/T<1$ and $(2\mu_e-Q)/T<1$,
Eq.~(\ref{eq:app_mue}) can be simplified further,
\be
Y_e=\frac{1}{2}\frac{(1+0.5Q/T)}{(1+1.5n_{\rmscr{B}}/T^3)}.
\label{eq:ye_app}
\ee
Equating the rates of $e^-$ and $e^+$ captures in the region of mild
degeneracy, \citet{2003ApJ...588..931B} obtained a
similar result to Eq.~(\ref{eq:ye_app})( see Eq.~(11) in his paper).
In a word, using the well known results and our 
new result Eq.~(\ref{eq:nd_eq}), we re-derive a similar 
previous result in a different way. 
From this consistence, it turns out that Eq.~(\ref{eq:nd_eq}) is 
correct and the approximate formula Eq.~(\ref{eq:app_mue}) is 
more accurate than the previous result.

\begin{figure}
\includegraphics[width=3.0in]{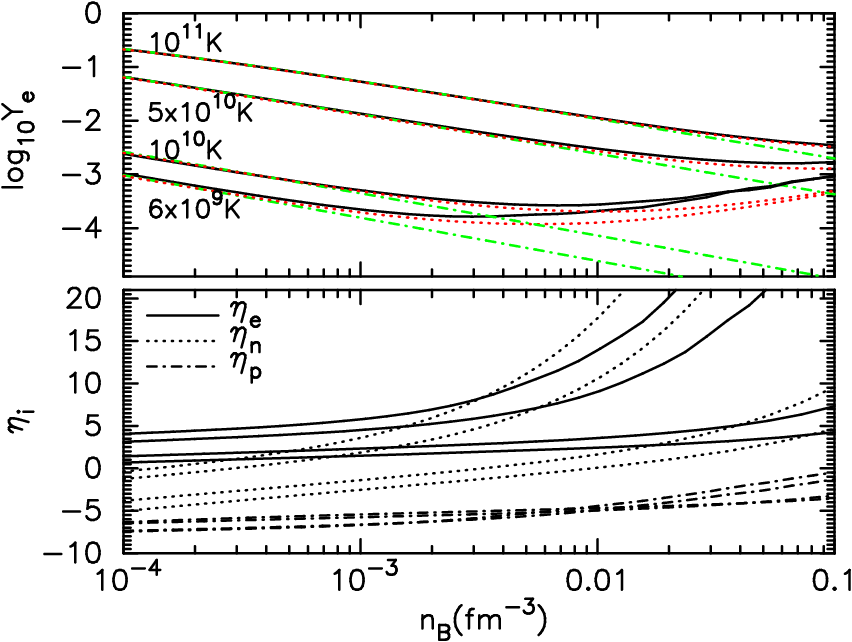}
\caption{
{\it Upper panel:} Electron fraction $Y_e$ under 
$\beta$-equilibrium
versus the baryon number density 
$n_{\rmscr{B}}$ at different temperatures. 
The solid curves show the results from 
Eqs.~(\ref{eq:rate_e_cap1})-(\ref{eq:rate_n_dec1}),
the dotted curves are from the analytic $\beta$-equilibrium 
condition Eq.~(\ref{eq:nd_eq}), and the dot-dashed curves are 
from the nondegenerate 
approximation of baryons.
{\it Lower panel:} The reduced chemical potentials of electrons, neutrons,
and protons $\eta_i$ as a function of the baryon number density at different 
temperatures. From top to bottom, the lines
correspond to the results of $T=6\times10^9$, $10^{10}$, $5\times10^{10}$, 
$10^{11}$K. \label{fig:frac}
}
\end{figure}

\begin{figure}
\includegraphics[width=3.0in]{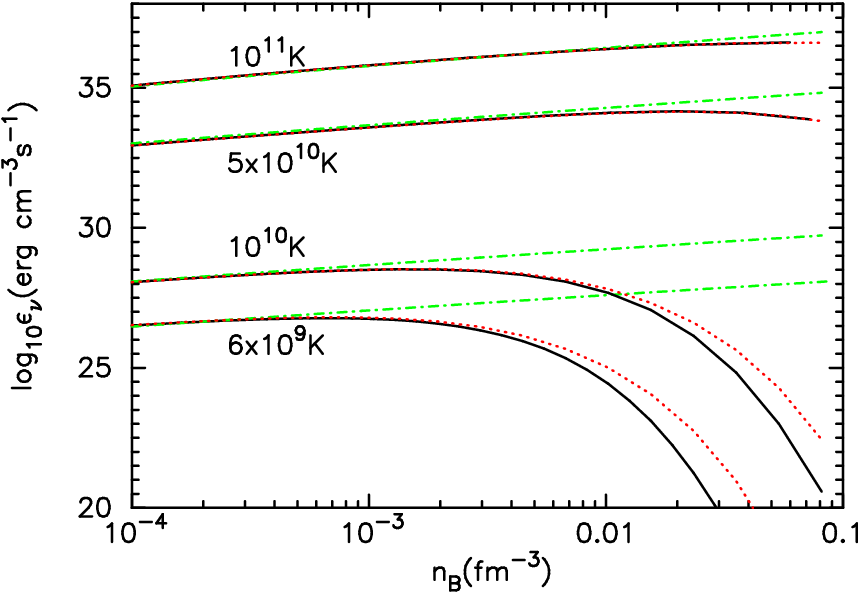}
\caption{
The neutrino emissivity $\epsilon_{\nu}$ versus the baryon number density at different 
temperatures. The lines are the same as those of the upper 
panel in Fig.~\ref{fig:frac}
\label{fig:emis}
}
\end{figure}

In the following, we check the validity of 
Eq.~(\ref{eq:nd_eq}) by numerical calculations.
As before, suppose that the nuclei are dissolved completely 
into nucleons at high temperature. Given $n_{\rm B}$ and $T$, the electron fraction 
$Y_e$ can be determined by two different sets of the equilibrium conditions. 
One set consists of 
Eqs.~(\ref{eq:equil})-(\ref{eq:charge}). 
In Eq.~(\ref{eq:equil}), the $\beta$ reaction rates are calculated 
based on the 
reaction rates Eqs.~(\ref{eq:rate_e_cap1})-(\ref{eq:rate_n_dec1}),
or the approximate rates Eqs.~(\ref{eq:rate_e_cap2})-(\ref{eq:rate_n_dec2}), 
respectively. 
The other set consists of our analytic equation Eq.~(\ref{eq:nd_eq}), and 
Eqs.~(\ref{eq:baryon})-(\ref{eq:charge}). 
Figure~\ref{fig:frac} shows $Y_e$ and the reduced chemical 
potentials $\eta_i$ ($i=n,p,e$)
versus the baryon number density at different temperatures. 
In the upper panel of Fig.~(\ref{fig:frac}), the solid lines show
the results from 
Eqs.~(\ref{eq:equil})-(\ref{eq:charge}) and
Eqs.~(\ref{eq:rate_e_cap1})-(\ref{eq:rate_n_dec1}) for the 
$\beta$ reaction rates. 
The dot-dashed curves are from 
Eqs.~(\ref{eq:equil})-(\ref{eq:charge}), but 
the approximate reaction rates in the nondegenerate limit 
Eqs.~(\ref{eq:rate_e_cap2})-(\ref{eq:rate_n_dec2}) are used.
The dotted lines represent the results from 
Eq.~(\ref{eq:nd_eq}) and Eqs.~(\ref{eq:baryon})-(\ref{eq:charge}).
It is evident that the results from the analytic condition 
are almost consistent with the general numerical results. As expected, 
the difference occurs only in the 
regime where electrons become degenerate, beyond which, the condition
for the analytic formula Eq.~(\ref{eq:nd_eq}) is not satisfied any
longer and the contribution from the neutron decay can not be neglected.
As shown in the lower panel of Fig.~\ref{fig:frac},
neutrons become degenerate
as electrons do. The particles become degenerate if their reduced 
chemical potential exceeds the temperature. For electrons and 
nucleons, their degenerate
number densities are estimated to be
\bea
n_e^{\rmscr{deg}}&=&p_e^3/3\pi^2\simeq T^3/3\pi^2 \nonumber \\
        &\simeq&9.0\times 10^{-6}\left(\frac{T}{10^{11}\rmmat{K}}\right)^3
        \rmmat{fm}^{-3}, \\
n_{n,p}^{\rmscr{deg}}&=&p_{n,p}^3/3\pi^2\simeq (2m_{n,p}T)^{3/2}/3\pi^2
        \nonumber \\
        &\simeq&2.8\times 10^{-3}\left(\frac{T}{10^{11}\rmmat{K}}\right)^{3/2}
        \rmmat{fm}^{-3}.
\eea
In the same regime,
there are great differences (of several orders) between our results and
the previous results
in which the degeneracy of nucleons is completely neglected.
In any case, Fig.~\ref{fig:frac} evidently shows that the results from the analytic condition 
are much more accurate than those from the approximate rates in all the 
parameter regions.

The total neutrino emissivity under $\beta$-equilibrium
is shown in Fig.~\ref{fig:emis}.
Compared with the general numerical results, both approximate 
methods overestimate the rate of the neutrino emission because 
neglect of the degeneracy of particles increases the phase space for 
the relevant reactions.
It is clearly shown in Fig.~\ref{fig:emis} that the effect of 
the neutron degeneracy 
is much more important than that of electrons, if such conditions are
satisfied.
 
\section{Conclusions and discussions}
In this work, 
using the structure function formalism developed by 
\citet{1998PhRvD..58a3009R}, 
we derive the rates of the $\beta$ processes which including
$e^{\pm}$ captures and neutron decay in the dense subnuclear 
matter for all participating particles are at arbitrary degeneracy.
For this purpose, the difference between the mass of 
neutron and proton is kept in the structure function 
(see Eq.~(\ref{eq:struct})), which is neglected originally in 
\citet{1998PhRvD..58a3009R}. Comparing to our reaction rates, the previous 
approximations have been checked. Especially, the electron capture
rate under the elastic approximation
which is put forward to include the effects of the 
degeneracy of baryons differs dramatically
from our results when baryons become degenerate 
and before the dense matter reaches the $\beta$-equilibrium.
Generally speaking, the elastic approximation underestimates 
the electron capture
rate when baryons become degenerate. 

Most of interest, we find an analytic steady state 
equilibrium condition for $e^{\pm}$
plasma with nucleons, if the system is transparent to the emitted neutrinos.
This result is valid when electrons are nondegenerate or mildly degenerate, 
but it has nothing to do with the degeneracy of baryons. 
Basically, it is a good result for $e^{\pm}$
plasma with nucleons, because when electron becomes degenerate, the 
number density of positron decreases exponentially, then the system 
can not be called as $e^{\pm}$ plasma. If the nucleons are nondegenerate, 
we further obtain only one simple equation for determining the 
electron chemical potential and
electron fraction, which shows the great advantage of our analytic 
equilibrium condition.

So far, there are four analytic beta equilibrium or steady state 
equilibrium conditions for
$npe^{\pm}$ gas in different astrophysical circumstances. They are 
summarized as follows:
(1), If the neutrinos are completely trapped in a system,
such as in the interior of a PNS,
$
\mu_{n}+\mu_{\nu_e} = \mu_{p} + \mu_{e}.
$
(2), In the cold $npe^-$ gas under beta equilibrium,
$
\mu_{n} = \mu_{p} + \mu_{e}.
$
(3), If the neutrinos are partially trapped, and the chemical potential
of the trapped neutrinos is zero ($\mu_{\nu_e}=0$),
$
\mu_{n}= \mu_{p} + \mu_{e}.
$
(4), If the neutrinos can escape freely from the system with plenty of 
e$^{\pm}$ pairs, 
$
\mu_{n} = \mu_{p} + 2\mu_{e},
$
which is the main result of this work.

\begin{acknowledgments}

The author would like to thank the anonymous referee for her/his
constructive suggestions which are very helpful to improve this 
manuscript, and Dr. Ramesh Narayan, Dr. Jeremy Heyl and 
Dr. Rosalba Perna for many discussions, Dr. Dong Lai for comments,
and Dr. David Rusin for a critical reading of this manuscript.
The author acknowledges the hospitality of Harvard-Smithsonian Center
for Astrophysics.  
This work is partially supported by the Special Funds for Major State
Research Projects, and the National Natural Science Foundation
(10233030).
\end{acknowledgments}

\end{document}